\documentclass{emulateapj}
\usepackage{graphicx}
\usepackage{natbib}
\usepackage{bm}
\shorttitle{Halo assembly bias}
\shortauthors{Dalal et al.}

\begin{document}
\title{Halo Assembly Bias in Hierarchical Structure Formation}
\author{Neal Dalal\altaffilmark{1}, Martin White\altaffilmark{2},
J.\ Richard Bond\altaffilmark{1} and Alexander Shirokov\altaffilmark{1}}
\altaffiltext{1}{Canadian Institute for Theoretical Astrophysics,
  University of Toronto, 60 St.\ George St., Toronto, ON, M5S 3H8, Canada}
\altaffiltext{2}{Department of Astronomy, University of California, 
Berkeley, CA 94720}

\begin{abstract}
We investigate the origin of halo assembly bias, the dependence of
halo clustering on assembly history.  We relate halo assembly
to peak properties measured in the Lagrangian space of the initial
linear Gaussian random density field, and show how these same Lagrangian 
properties determine large-scale bias.  We focus on the two regimes 
where assembly bias has been observed to be significant: at masses
very large and very small compared to the nonlinear mass scale.  At
high masses, we show that assembly bias is expected from the statistics
of the peaks of Gaussian random fluctuations, and we show that the
extent of assembly bias found in N-body simulations of rare halos is
in excellent agreement with our theoretical prediction.  At low masses,
we argue that assembly bias largely arises from a sub-population of
low mass halos whose mass accretion has ceased.  Due to their
arrested development, these halos naturally become unbiased, in
contrast to their anti-biased peers.  We show that a simple toy model
incorporating these effects can roughly reproduce the bias trends
found in N-body simulations.
\end{abstract}


\section{Introduction} \label{sec:introduction}

Hierarchical structure formation is a basic prediction of the highly
successful cold dark matter cosmological model, and a key element of
hierarchical growth is the formation of bound virialized
objects called dark matter halos. Dark matter halos are not
distributed uniformly throughout the universe, but rather compose 
a beaded filamentary network,  
sometimes referred to as the cosmic web \citep{BonKofPog96}.
The clustering of the rare, massive dark matter halos is enhanced
relative to the general mass distribution
\citep{Kai84,Efs88,ColKai89,BCEK,MoWhi96,SheTor99}, an effect known as
bias.  The more massive the halo, the larger the bias.  As a result,
the mass of halos hosting a given population of objects is sometimes
inferred by measuring their degree of clustering -- allowing a
statistical route to the notoriously difficult problem of measuring
masses of cosmological objects.  There are many applications for
statistical mass determination, including (for example) cluster
mass-observable self-calibration \citep[e.g.][]{LimaHu04,Wu08} or
reconstruction of the halo occupation distribution of galaxies or
quasars \citep[e.g.][]{CooShe02}.

Since halos of a given mass can differ in their formation history, properties
and large-scale environment\footnote{The large-scale environment of a halo
refers to the density, smoothed on some suitably large scale,
e.g.~$8\,h^{-1}$Mpc.},
a natural question arises: do these details affect halo clustering?
Numerical simulations are now able to produce samples with sufficient
statistics to test for the dependence of clustering on history, properties
and environment and recent numerical simulations reveal subtle dependences
of halo clustering on formation redshift, internal structure and spin
\citep{GaoSprWhi05,Wec06,Har06,Bet07,Wet07,JinSutMo07,GaoWhi07,AngBauLac08}

Using a marked correlation function, \citet{SheTor04b} found that
close halo pairs tend to have earlier formation times than more
distant pairs, work which was extended and confirmed by \citet{Har06}.
\citet{GaoSprWhi05} found that later forming, low-mass halos are less
clustered than typical halos of the same mass at the present.
\citet{Wec06} and \citet{Wet07} found a similar dependence upon halo
formation time (defined in a slightly different way), showing that the
trend reversed for more massive halos and that the clustering depended
on halo concentration.  Recently, \citet{GaoWhi07} showed that a wide
variety of halo properties correlate with bias, including spin and
substructure content, while \citet{LiMoGao08} showed that different 
definitions of halo age, when applied to the same halo samples, give
different amounts of assembly bias.

What remains unclear from this numerical work is the physical origin
of this effect.  \citet{GaoSprWhi05} argued that assembly bias is
inconsistent with the \citet{PreSch} model, which explicitly predicts
that halo abundance (and hence clustering) depends only upon halo
mass and no other internal properties related to assembly history.
Attempts to explain assembly bias based on the Press-Schechter
approach (i.e.\ the excursion set using a sharp $k$-space filter)
would therefore appear doomed from the outset, however several workers
have explored modifications to the Press-Schechter model in the hopes
of accounting for assembly bias.  \citet{San07} considered excursion
set mass functions using barrier heights determined from the
ellipsoidal collapse model of \citet{BonMye96}, but were unable to
account for the magnitude of assembly bias seen in simulations.
\citet{Mo05} suggested that ``pre-heating'', arising from the partial
collapse of structure into sheets and filaments, could suppress mass
accretion onto halos within those sheets and filaments.  This
mechanism would clearly be incapable of explaining assembly bias in
high-mass halos, which collapse {\em before} the formation of sheets
and filaments \citep{BonKofPog96}, and \citet{San07} found that this
mechanism fails in low-mass halos as well.  \citet{Des07} investigated
the effects of environment on the \citet{BonMye96} ellipsoidal
collapse model, but could not reproduce the sign reversal of assembly
bias between high and low masses, or the magnitude of assembly bias
seen in N-body simulations.

In this paper, we will argue that many aspects of assembly bias
may be understood quite simply.  In particular, we show that the
statistics of Gaussian random fields require significant assembly bias
in high-mass halos, and we show using N-body simulations that the
clustering of massive halos matches precisely the assembly bias that
we predict.  At low halo masses, we argue that assembly bias arises
largely due to the cessation of mass accretion onto a sub-population
of small halos.  These non-accreting halos then naturally become
unbiased over time, and therefore cluster more strongly than the
majority of halos of similar mass, which are anti-biased.

\section{Bias} \label{sec:bias}

In this section, we briefly review previous results on biasing and
assembly bias.  

We start by assuming that massive dark matter halos arise from the
collapse of peaks in the initial (linear) density field; this allows
us to infer many of the properties of halo statistics using the
statistics of gaussian fields \citep{PreSch,BBKS,BCEK,BonMye96}.  
High peaks of the density field will in general cluster more strongly than
the average matter clustering, which leads to halo bias.  In the
linear bias model, we directly relate the halo fluctuations to the
local matter density fluctuations,
\begin{equation}
\delta_h = b\,\delta,
\label{linearbias}
\end{equation}
where $\delta\equiv\delta\rho/{\bar\rho}$ is the overdensity and $b$
is the bias parameter.  On scales over which this expression is 
valid\footnote{This approximation should be valid on scales where the
matter correlation is small, $\xi(r)\ll 1$
or $\Delta^2(k)=k^3P(k)/2\pi^2\ll 1$ \citep{Kai84}.  In general the
scale-dependence of the bias will extend to larger scales the more
biased the halos.},
we may express halo two-point functions in terms of the matter
two-point function, scaled by appropriate powers of the bias.  For
example, the halo power spectrum would become $P_h(k)=b^2 P_m(k)$.

The clustering of halos reflects the clustering of initial peaks only
in part, however.  Another contribution to halo clustering comes from
the motion of peaks along large-scale matter flows.  At late times,
the halo overdensity becomes
\begin{equation}
\delta_h(a) = \delta_{\rm L} + \delta_m(a)
\end{equation}
where the Lagrangian overdensity $\delta_{\rm L}$ corresponds to the
clustering of the peaks in the initial density field, and the second
term $\delta_m$ reflects the motion of the peaks due to advection
along matter flows.  On sufficiently
large scales, in which we can treat the motion of the peaks as tracing
the average bulk motion, the second term is simply the matter
overdensity. Then the (Eulerian) bias becomes
\begin{eqnarray}
b(a) &=& \delta_h(a)/\delta_m(a) = 1 + \delta_{\rm L}/\delta_m(a)
\nonumber\\
&=& 1 + b_{\rm L}(a).
\label{eulerian}
\end{eqnarray}
Note that we explicitly include the time dependence of the matter
overdensity, but that the Lagrangian overdensity of peaks is not time
dependent: it simply reflects the clustering of peaks in the initial
conditions.  Therefore, over time, the clustering associated with
motion will dominate over the clustering from the initial conditions.
That is, if we can treat the halos as test particles, we expect their
Lagrangian bias to decay with time inversely with the linear growth
factor, $b_{\rm L}(a) \propto 1/D(a)$ \citep[e.g.][]{Fry96,TegPee98}.  

Henceforth, we will focus on the Lagrangian bias $b_{\rm L}$, noting
that the conversion to Eulerian bias $b$ is simply given by
Eqn.~(\ref{eulerian}) on linear scales.  The next step is to determine
$b_{\rm L}$. From Eqn.~(\ref{linearbias}), we can think of bias as
the rate at which the number of halos or peaks changes as the background
density is varied.  If we associate dark matter halos
with rare peaks above some high threshold density, then it becomes
easy to see why massive halos are biased.  In the Press-Schechter
model, for example, the (Lagrangian) number density of halos is 
\begin{equation}
n_{\rm PS} \propto \frac{\delta_c}{\sigma^2}
\exp\left(-\frac{\delta_c^2}{2\sigma^2}\right),
\end{equation}
and changing the background density level by $\delta$ is equivalent to
changing the threshold density to $\delta_c-\delta$.  Then the bias 
becomes 
\begin{equation}
b_{\rm PS}=n^{-1}\frac{dn}{d\delta}=-n^{-1}\frac{dn}{d\delta_c}
=\frac{\delta_c}{\sigma^2} - \frac{1}{\delta_c}
\end{equation}
\citep{ColKai89}.  This precise expression applies only for the
Press-Schechter mass function, but its qualitative features are
generally valid. At high masses $(\sigma\ll\delta_c)$, peaks above
threshold are much more clustered than matter, because changing the
background density level dramatically changes the probability out on the
tail of the Gaussian distribution.  In contrast, at low masses
$(\sigma\gg\delta_c)$, halos are anti-biased ($b_{\rm L}<0$) because
raising the background density level reduces the number of peaks at
low mass, by converting them to higher mass.

Clearly, many aspects of the mass dependence of bias may be inferred
by consideration of the initial density peaks from which halos form.
We can largely understand the clustering of Eulerian halos
found in complex nonlinear simulations by adopting a Lagrangian
viewpoint, and examining the properties of peaks of the initial linear
density field.  
As mentioned in the introduction, simulations have shown that bias
depends upon additional halo properties besides mass, such as assembly
history or structural parameters like concentration.  The philosophy
of this paper will be to assume that this additional dependence may
also be understood from the statistics of the initial Gaussian peaks.
Accordingly, we will study N-body halos found in simulations described
below, using a Lagrangian viewpoint.  

\section{Simulation and post-processing} \label{sec:simulation}

We have performed N-body simulations of structure formation in
scale-free cosmologies, with $\Omega_m=1$ and power-law power
spectra.  We have chosen to run scale-free models rather than more
realistic $\Lambda$CDM models in order to simplify the problem as much
as possible, and remove possible effects from late-time cosmic
acceleration or a varying spectral index. 
On scales far from the box size and far from the Nyquist
frequency, the only relevant scale in the problem is $M_\star(a)$, the
characteristic nonlinear mass scale satisfying
$\sigma(M_\star,a)=\delta_c=1.68$.  If the power spectrum slope is
$n$, such that $P(k)\propto k^n$, then the rms mass fluctuations scale
like $\sigma(M)\propto M^{-(3+n)/6}$.  Because the linear growth
factor behaves as $D(a)=a$ for this critical-density cosmology, the
nonlinear mass scales like $M_\star(a) \propto a^{6/(3+n)}$.  Since
the nonlinear mass varies rapidly with expansion factor $a$, by
examining halo properties at multiple redshifts we can study halo
properties over a wide range of $M/M_\star$ \citep{Efs88,Wec06}. 

In this paper, we will focus on the largest simulation we ran, with
$1024^3$ particles, using a power spectral index $n=-2$.  We
normalized the power spectrum such that $\sigma(M,a)=10 a M^{-1/6}$,
where the mass $M$ is measured in number of particles.  The simulation
was evolved using the adaptive P$^3$M code {\tt gracos}
\citep{gracos}, using a Plummer softening length $\varepsilon_P=0.1$
times the mean interparticle spacing.  The simulation was started at
$a=0.005$, and 100 outputs were generated,
spaced evenly in $\log a$ from $a=0.0825$ to $a=1$, containing
phase-space coordinates (comoving positions and peculiar velocities)
and ID's for each particle.

The phase space data for each output is used to find the (sub-)halos in a
two-step process.  First we generate a catalog of halos using the
Friends-of-Friends (FoF) algorithm \citep{DEFW} with a linking length of
$b=0.2$ 
times the mean inter-particle spacing.
This procedure partitions the particles into equivalence classes by linking
together all particles separated by less than a distance $b$, with a density
of roughly $\rho>3/(2\pi b^3)\simeq 100$ times the background density.
We keep all halos with more than 32 particles.

We identify subhalos of the FOF groups using a new
implementation of the {\sl Subfind\/} method of \citet{Spr01}
which defines subhalos as gravitationally self-bound aggregations of
particles bounded by a density saddle point.  After some experimentation
with different techniques we found this method gave a good match to what
would be selected ``by eye'' as subhalos.
We use a spline kernel with 16  neighbors to estimate the
density and keep all subhalos with more than 20 particles.
We call the most massive subhalo in any FoF group the central halo.
The other subhalos we refer to as satellites.

For each subhalo we compute and store a number of properties including the
bound mass, velocity dispersion, peak circular velocity ($v_c$), total
potential energy, position and velocity.
Since mass loss can be quite extreme for some subhalos we will use $v_c$
rather than mass to quote subhalo `size'.  The mass, peak circular velocity
and potential are highly correlated, however since our subhalos are
not spherical, there is a non-trivial scatter between $v_c$ and mass at fixed
time.

A merger tree is computed from the set of subhalo catalogs by identifying
for each subhalo a ``child'' at a later time.  We process 4 consecutive
simulation outputs at a time and for each subhalo in the earliest output
we define its child as the subhalo at a later time which maximizes
\begin{equation}
  \alpha = \ln^{-1}\left({a_2\over a_1}\right)
    \left[ 1 - {\left|M_1-M_2\right|\over M_1+M_2}\right]
    \sum_{i\in 2} \phi_{1i}^2
\end{equation}
where $a_1$ and $M_1$ are the scale-factor and mass of the progenitor,
$a_2$ and $M_2$ are the candidate halo at the later time, $\phi_{1i}$ is
the potential of particle $i$ computed using the particles in subhalo 1
and the sum is over all particles in the progenitor that also lie in the
candidate.
The weighting by $\phi^2$ ensures that the ``core'' particles in the
progenitor are given more weight than the less bound particles.  We found
that weighting by $\phi^2$ gave better results than weighting by $\phi$,
but higher powers of $\phi$ did not perform appreciably better than $\phi^2$.
We find $\sim 95\%$ of the subhalos have a child in the next time step, with
one or two percent skipping one or more output times.  A similar
method for subhalo tracking has been employed by \citet{Millenium},
\citet{FAGYH}, \citet{Krav04}, \citet{All06}, and \citet{Har06}.

In addition to the Eulerian halo properties mentioned above, we also
require Lagrangian properties of halos, such as their centroids or
mean initial velocities.  From the initial conditions for the
simulation, we have the initial density field $\delta({\bm x})$, the
initial comoving displacement field ${\bm d}({\bm x})$ satisfying
$\delta+\nabla\cdot{\bm d}=0$, the peculiar velocity field, which in
the Zeldovich approximation for $\Omega_m=1$ is ${\bm v}=aH\,{\bm d}$,
and the strain field ${\bf S}=\nabla{\bm d}$.  Note that ${\rm Tr\
}{\bf S}=-\delta$.  These fields are all computed on the same grid as
the initial (undisplaced) particle positions, by Fourier transform of
the initial density field.  For example, the strain field $S_{ij}$ is
computed by FFT of $\delta_{\bm k} k_ik_j/k^2$.  
We would like to average these quantities
over the Lagrangian volumes occupied by the halos.  We may do so using
the unique ID's for each particle.  For each field $F$, we compute the
average $\langle F\rangle$ over every halo by a simple average over
the halo's particles:
\begin{equation}
\langle F\rangle_i = N_i^{-1} \sum_{j=1}^{N_i} F({\bm x}_{i,j}),
\end{equation}
where ${\bm x}_{i,j}$ is the Lagrangian position of the $j^{\rm th}$
particle of halo $i$.  Lastly, we will also require derivatives of
these smoothed properties with respect to mass, e.g.\
$d\langle\delta\rangle/d(\log M)$.  We use the merger tree described
above to compute derivatives with respect to mass.  For every halo, we
move along the most massive branch of its merger tree for a factor of
2 in mass, tabulating mass $M$ and the average
$\langle F\rangle$.  We then fit a linear expansion, $F=F_0 + F^\prime
\Delta(\log M)$ in order to measure the derivative $F^\prime\equiv
dF/d\log M$.  

Lastly, we will determine bias parameters in Fourier space.  We
measure the matter power spectrum $P(k)$, the halo auto-power spectrum
$P_h(k)$, and the halo-matter cross spectrum $P_c(k)$, and define the
halo bias as $b=P_c(k)/P(k)$.  We use the ratio of the
cross-spectrum rather than the halo auto-spectrum, because the former
is less sensitive to shot noise in the halo counts. 

\section{High mass halos: $M\gg M_\star$} \label{sec:highmass}

\begin{figure}
\centerline{\includegraphics[width=0.5\textwidth]{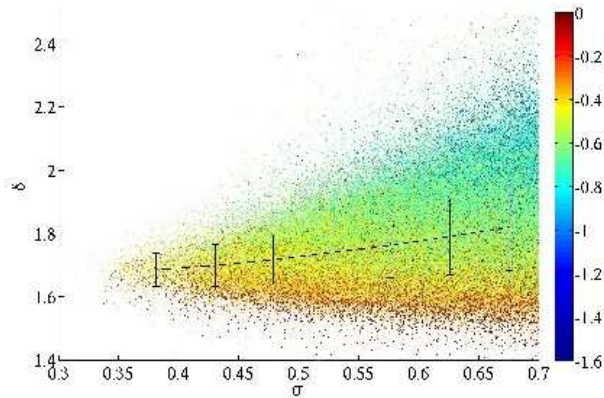}}
\caption{Average Lagrangian overdensity $\langle\delta\rangle$ for
  halos as a function of $\sigma(M)$.  In the high mass limit,
  $\sigma\to 0$, collapse becomes spherical and
  $\langle\delta\rangle\to\delta_c=1.68$.  Point colors correspond to
  peak curvature, $d\langle\delta\rangle/d\log M$, as denoted in the
  colorbar on the right.  The solid black curve and error bars depict
  the mean and standard deviation of $\delta$ in bins of $\sigma$.  
\label{sigmadelta}
}
\end{figure}

We first focus on halos well above the nonlinear mass scale, $M\gg
M_\star$, where $\sigma(M)\ll\delta_c$.  These halos are particularly
simple : because they are so massive, they dominate their
environments, and because they are far out on the tail of the
probability distribution of $\delta$, the mass dependence of their
bias scales like $b \simeq \delta_c/\sigma^2$.  Anti-biasing from
larger-scale halos should be negligible.  Another key
simplification in this regime is that the collapse of these rare peaks
is nearly spherical \citep[Appendix C]{BBKS}.
Because the collapse is roughly spherical, we expect the collapse
threshold predicted by the spherical collapse model \citep{GunGot72}
to be a good approximation: $\delta_c \approx 1.68$.  This expectation
is borne out by our N-body results.  In figure \ref{sigmadelta}, we
plot the average linearly evolved overdensity for FOF halos well above
$M_\star$, and indeed we find $\langle\delta\rangle\simeq 1.7$ for
the $\gtrsim 3\sigma$ peaks. Note that multiple
different redshifts and halo masses are plotted in this figure.

The validity of $\delta_c\approx 1.68$ for halos in the regime
$\sigma\ll 1$ allows us easily to estimate the halo bias as
$b\sim\delta_c/\sigma^2$ as mentioned above.  This is the average over
all peaks of a given mass, which marginalizes over any other peak
parameters which could affect bias.  One obvious additional parameter
which the bias must depend upon is the curvature of the peak, which we
will parametrize by $s=d\langle\delta\rangle/d(\log M)$.  To see this,
note that we can estimate the large-scale background density
$\delta_b$ in which the halos reside by a Taylor expansion: $\delta_b
\approx \delta + \Delta(\log M)\times d\delta/d(\log M)+\ldots$.  Two
peaks of the same mass $M$ and peak height $\delta$ but different peak
curvatures $s_1 > s_2$ (i.e.\ $|s_1| < |s_2|$)
will tend to live in different environments:
peak 1 will tend to have a larger background density than peak 2, and
therefore will have a larger bias.  It is straightforward to estimate
the dependence of bias on curvature.  Writing $\nu=\delta/\sigma$,
$x=s/\sigma_s$, and the cross-correlation coefficient
$\gamma=\langle\nu x\rangle$, the bias becomes
\citep{BBKS}\footnote{\citet{BBKS} parametrized curvature not with
  $d\delta/d\log M$ as we do, but rather with the closely related
  $\nabla^2\delta$.}
\begin{equation}
b_{\rm L} \approx \sigma^{-1}\frac{\nu-\gamma x}{1-\gamma^2}.
\label{bias_curv}
\end{equation}
We can easily sketch a derivation of this expression following the
argument of \citet{Kai84}.  Let us compute the correlation function
for regions above a high threshold $\nu_t$.  Consider regions 1 and 2,
each of size $R$ and separated by distance $r_{12}$, with peak heights
$\nu_1,\nu_2$ and curvatures $x_1,x_2$.  These Gaussian fields are
described by a covariance matrix with diagonal elements
$\langle\nu^2\rangle=\langle x^2\rangle=1$.  Of the off-diagonal
elements, only $\langle\nu_1\nu_2\rangle=\psi(r_{12})$ and
$\langle\nu_1x_1\rangle=\langle\nu_2x_2\rangle=\gamma$ are important;
long-range correlations involving derivatives of the density field
fall off more quickly than these terms by powers of $R/r_{12}$.  We
compute the two-point correlation of regions with $\nu>\nu_t$ by
integrating the Gaussian probability over $\nu_1>\nu_t$,
$\nu_2>\nu_t$.  If we are uninterested in the curvature $x$, then we
first marginalize over $x_1$ and $x_2$, leaving a $2\times 2$
covariance matrix with off-diagonal element $\psi$, and we find that
the two-point function for regions above threshold becomes 
$\xi_{\rm pk}=\nu_t^2\psi$, giving a Lagrangian bias 
$b_{\rm L}=\nu_t/\sigma$ \citep{Kai84}.  We can similarly compute the
two-point function for regions above threshold with a specified
curvature $x$ by repeating this calculation, but not marginalizing
over $x_1$ and $x_2$.  Recalling that the one-point probability
distribution for $\nu$ at a given $x$ is a Gaussian centered at
$\langle\nu|x\rangle=\gamma x$ with variance $1-\gamma^2$, we change
variables from $\nu$ to 
$\nu^\prime=(\nu-\gamma x)/(1-\gamma^2)^{1/2}$.  Noting that
$\langle\nu_1^\prime\nu_2^\prime\rangle\approx\psi/(1-\gamma^2)$, we
immediately see that the two-point correlation function for regions
above threshold becomes 
$\xi_{\rm pk}=(\nu_t^\prime)^2\psi/(1-\gamma^2)$, giving a bias 
$b_{\rm L}=\nu_t^\prime/(\sigma \sqrt{1-\gamma^2})$, i.e.\
Eqn.~(\ref{bias_curv}).

\begin{figure}
\centerline{\includegraphics[width=0.4\textwidth]{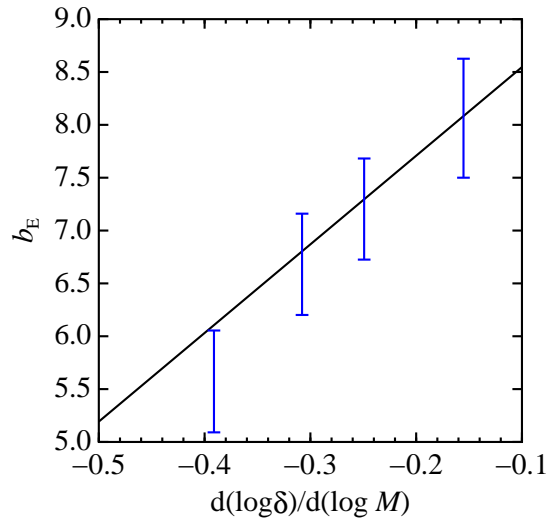}}
\caption{Eulerian bias $b_{\rm E}=1+b_{\rm L}$ for halos with 110-130
  particles at $a=0.1089$, as a function of peak curvature. These
  halos have $\sigma=0.49$, and are roughly $\sim 1600 M_\star$ at
  this redshift.  Halos are binned into quartiles of
  $d\log\delta/d\log M$, and halo bias in each bin is measured by the
  ratio of the halo-matter cross-spectrum to the matter power
  spectrum.  The solid black curve shows Eqn.~(\ref{bias_curv}).
\label{high_bias}
}
\end{figure}

The bias of halos found in our simulation matches
Eqn.~(\ref{bias_curv}) quite well.  For a power spectrum index $n=-2$,
as we have used, note that $\sigma_s=\sigma/\sqrt{6}$ and
$\gamma=-1/\sqrt{6}$, for a top-hat window function.  In figure
\ref{high_bias} we plot the halo bias as a function of peak curvature,
along with our prediction from Eqn.~(\ref{bias_curv}).  There is
evidently a strong dependence of bias upon peak curvature.  The $\sim
40\%$ range in $b$ corresponds to a factor of 2 variation in the
correlation function for halos of this mass.

\begin{figure}
\centerline{\includegraphics[width=0.4\textwidth]{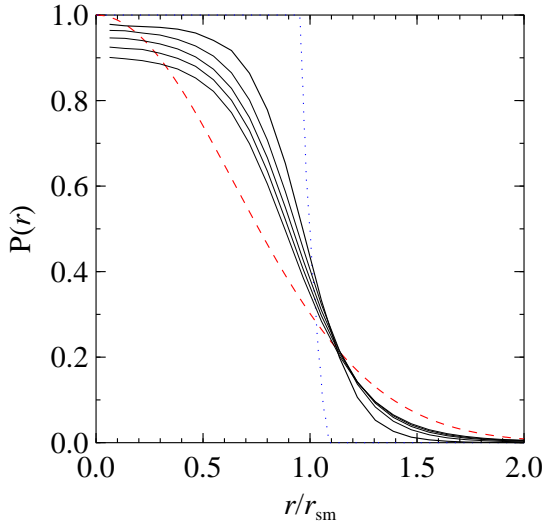}}
\caption{Probability for a particle to be linked into a high mass FOF
  group as a function of initial Lagrangian radius.  The thick solid
  black curve 
  shows the stacked Lagrangian volumes for halos in the mass range
  200-300 particles at $a=0.1089$, corresponding to roughly $\sim 3000
  M_\star$.  Note that mass in subhalos is excluded from this average.
  The blue dotted curve shows the expected
  average profile for a tophat window, while the red dashed curve
  corresponds to a Gaussian of this mass range.  The thin solid black
  curves show the average Lagrangian profiles for halos of mass
  between 200-300 particles at later outputs,
  $a=0.1896, 0.3300, 0.5547$, and 1.0000, corresponding to $M/M_\star
  \simeq 100, 4, 0.2$ and 0.005 respectively.
\label{stack_vol}
}
\end{figure}

One subtlety which we have glossed over in computing the statistics
of high-mass peaks is the issue of the appropriate smoothing to use. 
We have assumed a top-hat smoothing, however there
is no compelling theoretical reason to favor a top-hat kernel versus
other possibilities, such as a Gaussian (although the sharp $k$-space
filter implicit in the Press-Schechter model appears unphysical).  As
a sanity check on our top-hat smoothing, we have computed the average
Lagrangian volumes corresponding to high peaks, by stacking roughly
1000 halos of mass $\sim 3000 M_\star$.  The results are shown in
figure \ref{stack_vol}.  The average Lagrangian volume occupied by FOF
halos is reasonably compact, with the probability for halo inclusion
steeply varying from near 1 to near 0 over a range $0.8<r_{\rm TH}<1.2$, 
where $r_{\rm TH}$ is the top-hat smoothing radius for mass $M$.  In
comparison, a Gaussian smoothing window is clearly more diffuse.  We
conclude that our use of top-hat smoothing is reasonable, though we
caution that these results may depend upon the slope of the power
spectrum.  Finally, we note that there is a nontrivial amount of mass
inside $r<r_{\rm TH}$ that does not end up in the halo; we will return
to this point below in section \ref{sec:lowmass}.

\subsection{Halo observables}

We have shown that there is a strong dependence of halo clustering
upon peak curvature.  How does this relate to the assembly bias
observed in previous simulations?  

\begin{figure}
\centerline{\includegraphics[width=0.4\textwidth]{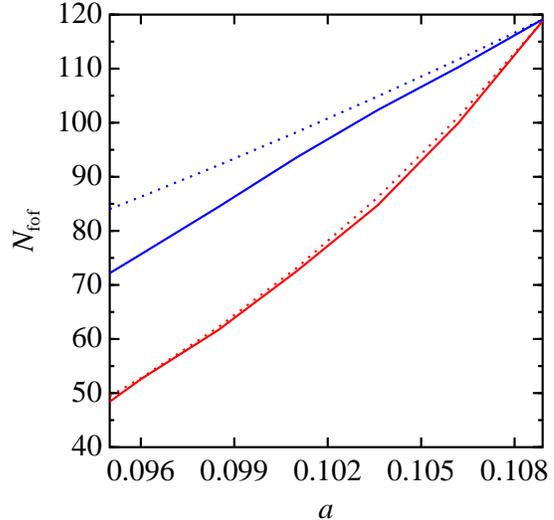}}
\caption{Mass accretion histories as a function of peak curvature.
  The upper solid (blue) curve shows the average accretion history for
  the halos comprising the leftmost bin in Fig.~\ref{high_bias}, while
  the lower solid (red) curve shows the average history for the
  rightmost bin.  The dotted curves depict the behavior $M\propto
  a^\alpha$ for $\alpha=-1/(d\log\delta/d\log M)$ expected from the
  spherical collapse model.
\label{history}
}
\end{figure}

In the high mass regime, there is a direct relationship between peak
curvature and halo assembly history.  Recall that in the spherical
collapse model, regions complete their collapse once the average
linearly evolved overdensity $\delta$ reaches $\delta_c$.  This implies
that we can estimate the assembly history of halos simply by measuring
how the smoothed overdensity $\langle\delta\rangle$ varies with
smoothing scale.  We would expect the accretion rate and the curvature
to be related by $d(\log M)/d(\log a) = -[d(\log\delta)/d(\log M)]^{-1}$.  
The average mass accretion histories for high mass halos appears
consistent with this expectation.  In figure \ref{history} we plot the stacked
mass accretion history profiles for halos of 120 particles at $a=0.1089$,
splitting the population based on peak curvature.  The average
accretion rates are in good agreement with our expectations.  This is
true only in the average, however : individual accretion histories
are noisy because much of the accretion onto halos
is not smooth, but rather consists of discrete accretion events.  This
may explain why different age indicators for halos give different
levels of assembly bias \citep{Wet07}.  The noisiness of commonly used
age indicators may be large enough to mask the strong assembly bias
clearly present in high mass halos.

\begin{figure}
\centerline{\includegraphics[width=0.4\textwidth]{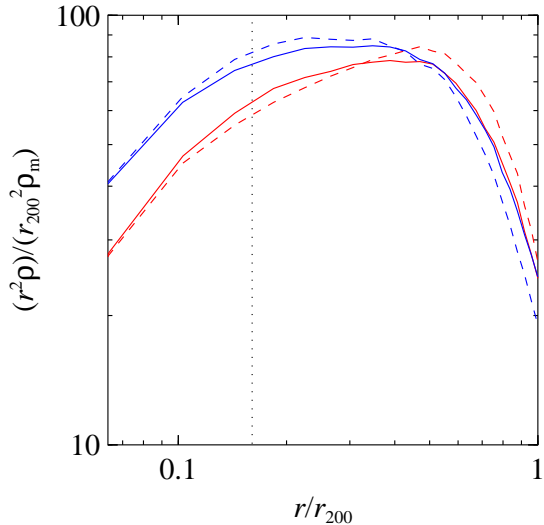}}
\caption{Radial profiles for halos as a function of peak curvature.  
  The various curves show the stacked radial profiles for subsets of
  halos of 4000-5000 particles at $a=0.2501$, roughly $M\sim 400
  M_\star$.  The solid curves correspond to the upper and lower
  quartiles of $d\log\delta/d\log M$, while the dashed curves
  correspond to the upper and lower quartiles of fitted concentration
  $c_{200}$.  For comparison, the vertical dotted black line shows
  $r=2.8\varepsilon_P$, the softening length.
\label{conc}
}
\end{figure}

Another halo parameter seen to correlate with bias is concentration
\citep{Wec06}.  While we cannot write down an expression directly
relating peak curvature and halo concentration, nevertheless we would
expect curvature and concentration to be correlated.  Previous work
has shown that assembly history and concentration are correlated
\citep{Wec02,Zhao03a,Zhao03b}, so the relation between peak curvature
and accretion rate discussed above suggests that concentration and
peak curvature may be correlated.  Physically, we would expect that
highly curved peaks should produce more concentrated halos.  To test
this, we plot in figure \ref{conc} the stacked radial profiles of
halos with 4000-5000 particles at $a=0.2501$, as a function of
Lagrangian peak curvature.  There is a correlation between initial
peak curvature and final halo concentration apparent in the stacked
profiles, although it is quite apparent that peak curvature is not
equivalent to concentration.  While the inner profiles (at $r<r_s$) of
curvature-selected halos match well with profiles of
concentration-selected halos, the outer profiles (at $r>r_s$) begin to
diverge.

Within the statistical errors of our finite sample of rare halos, we
have found no residual dependence of bias on concentration, once we
account for the peak curvature dependence described above.  Our
simulation had relatively low force resolution, and so our fitted
concentrations may be sufficiently noisy to mask any underlying
residual concentration dependence of bias, but it appears that
assembly bias in rare, high mass halos is in
agreement with theoretical expectations.  The interpretation of this
effect is straightforward: halos residing in high density environments
have an enhanced mass supply, and therefore grow faster, than halos in
low density environments.

\section{Low mass halos: $M\ll M_\star$} \label{sec:lowmass}

In this section we consider assembly bias for low mass halos with
$M\ll M_\star$, for which $\sigma(M)\gg\delta_c$.  The age- and
concentration-dependence of bias at low masses is considerably
stronger than in the high mass regime, and of the opposite sign: the oldest,
highly concentrated halos have a bias more than twice as large as that
of the youngest, low-concentration halos of similar mass $M\ll M_\star$
\citep{GaoSprWhi05,Wec06,GaoWhi07}.

\begin{figure}
\centerline{\includegraphics[width=0.4\textwidth]{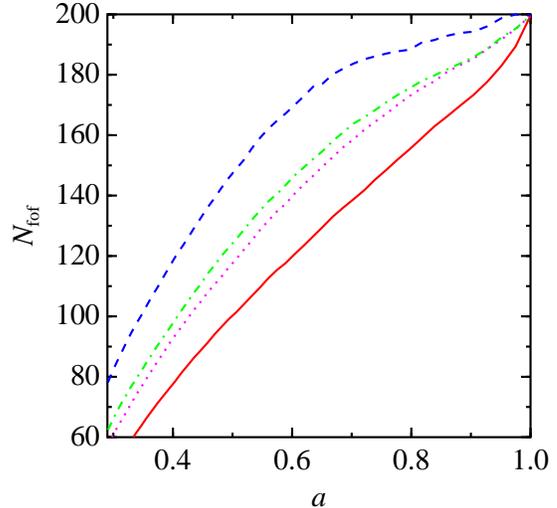}}
\caption{Mass accretion histories for halos of 190-210 particles
  at $a=1$, corresponding to $M=0.0045 M_\star$.  The halos are split
  into quartiles of $\langle\delta\rangle$, which correlates well with
  mass accretion history.
\label{history_lowm}
}
\end{figure}

In agreement with previous workers, we also find this reversed sense
of assembly bias.  In keeping with our Lagrangian viewpoint, we use
mean Lagrangian overdensity $\langle\delta\rangle$ as a proxy for
age. In figure \ref{history_lowm}, we plot average accretion histories
for halos of 200 particles at $a=1$, split into quartiles of
$\langle\delta\rangle$.  The oldest halos show little growth at late
times, accreting only $\sim 20\%$ of their mass after $a=0.5$.  In
comparison, the nonlinear mass scale increases by a factor $\approx
64$ over this same time.

\begin{figure}
\centerline{\includegraphics[width=0.4\textwidth]{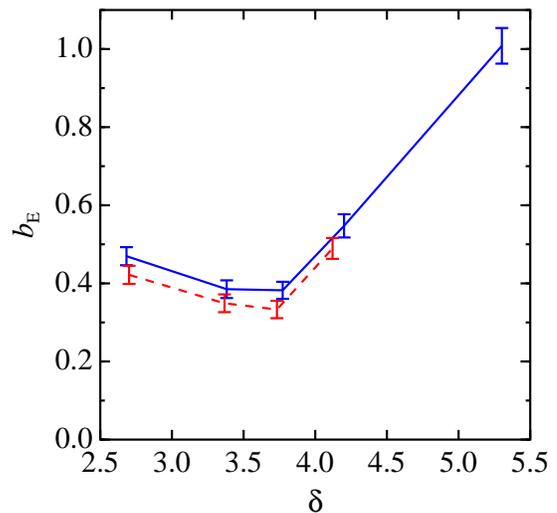}}
\caption{Eulerian bias for the same halos of $\sim200$ particles shown
  in Figure \ref{history_lowm}.  The oldest halos, at high
  $\langle\delta\rangle$, are nearly unbiased while the rest of the
  population is strongly antibiased.  Removing all halos which
  previously were subhalos, and all halos bound to more massive halos,
  removes the old unbiased subpopulation (red dashed curve). 
\label{bias_lowm} 
}
\end{figure}

Because the oldest halos do not grow appreciably in mass, we can think
of their dynamics as those of test particles following large-scale
bulk flows.  As discussed in section \ref{sec:bias}, we therefore
expect the bias of this population to relax over time to $b\to 1$.
This is indeed the behavior observed for our oldest halos, as shown in
figure \ref{bias_lowm}, which plots the (Eulerian) bias of the same
halos shown in figure \ref{history_lowm}. The oldest $\sim 20\%$ of
halos are nearly unbiased, while the remainder are strongly
anti-biased, to a degree consistent with predictions of even the
simplest Press-Schechter model, $b\to 1-\delta_c^{-1}\approx 0.4$
\citep{ColKai89}.

The magnitude of assembly bias present at low masses is therefore not
surprising.  Most low-mass halos are anti-biased at expected levels,
while a subpopulation becomes unbiased.  The theoretical challenge is
to explain the number of unbiased, low-mass halos.

One hint towards the origin of the non-accreting, unbiased population
comes from the observation that many, if not most of these halos are
associated with nearby larger halos \citep[e.g.][]{Wang07}.
In figure \ref{bias_lowm}, we
again plot the bias of $\sim 200$ particle halos, but excluding those
which in earlier outputs were subhalos. This removes $\sim 20-25\%$ of
the total population.  Because the time sampling of our simulation
outputs do not finely cover the dynamical times of all halos, this cut
could miss subhalos with orbital pericenters inside of larger halos,
but apocenters well outside.  \citet{Lud08} have recently argued that
a significant fraction of subhalos have orbital apocenters extending
out to many times the virial radius of their host halo.  To account
for this, we also conservatively exclude all halos which are within 3
$r_{\rm vir}$ and gravitationally bound to larger halos.  This will
exclude the population discussed by \citet{Lud08}, as well as any
halos entering the infall region of massive objects.  This cut
removes a further $\sim 15\%$ of low mass halos, leaving $\gtrsim 60\%$
of the original low-mass population.  The effect of these cuts is
effectively to exclude the old subpopulation, even though the cuts do
not explicitly refer to halo age or accretion history.  It therefore
does seem likely that, as suggested by previous authors, the old
subpopulation of low-mass halos corresponds to early-formers whose
growth was stunted by environmental effects \citep{Wang07}.

\begin{figure}
\centerline{\includegraphics[width=0.4\textwidth]{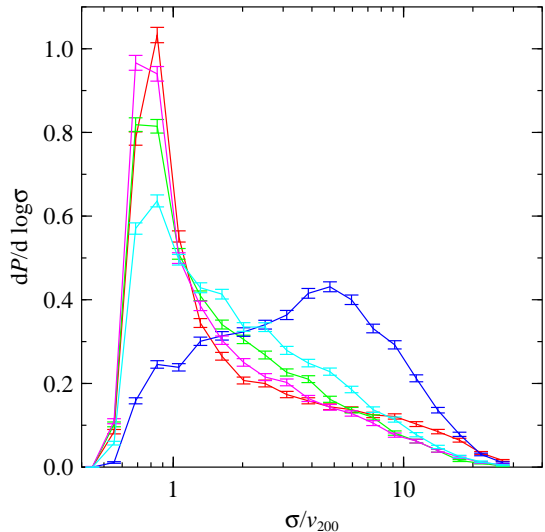}}
\caption{Probability distribution for ambient 3-D velocity dispersion
  $\sigma$ within a radius $5 r_{200}$ of halos with
  200-300 particles at $a=1$, roughly 0.005 $M_\star$. Halos are split
  into 5 bins of $\langle\delta\rangle$, with  red, magenta, green,
  cyan, and blue from lowest to highest respectively.  The oldest
  halos tend to live in hot environments, whereas the material in the
  vicinity of most low-mass halos is sufficiently cold to permit
  accretion. 
\label{sigv}
}
\end{figure}

One plausible environmental effect that would account for the
arrested development of the oldest low-mass halos is simply the
enhanced velocity dispersion in the vicinity of massive halos.  In
environments where the ambient velocity dispersion greatly exceeds the
escape velocity of low-mass halos, those halos will be unable to
accrete significant material.  We indeed find that the oldest halos
preferentially reside within hot environments, as shown in
Fig.~\ref{sigv}. 

Given that the oldest low-mass halos tend to exist in proximity to
larger halos, we would expect them to originate (in Lagrangian space)
from the vicinities of high mass halos as well, c.f.\
Fig.~\ref{stack_vol}.  As discussed in the previous section, much of 
the material within the Lagrangian volumes of high mass halos ends up
within the final collapsed halo, so how do we understand the survival
of the old low-mass halos in the presence of their rapidly accreting
neighbors?  One possible explanation noted by \citet{Lud08} is that
these halos are ejected out of larger halos by 3-body interactions.  
Because our goal is to understand the
clustering of halos in terms of their Lagrangian properties of known
(Gaussian) statistics, and since complicated multi-body interactions are
clearly difficult to model using only Lagrangian quantities, we have
not attempted to address this mechanism using our simulations.
Rather, we have tried to find Lagrangian properties of halos which
correlate with their survival around larger structures. 

\begin{figure}
\centerline{\includegraphics[width=0.45\textwidth]{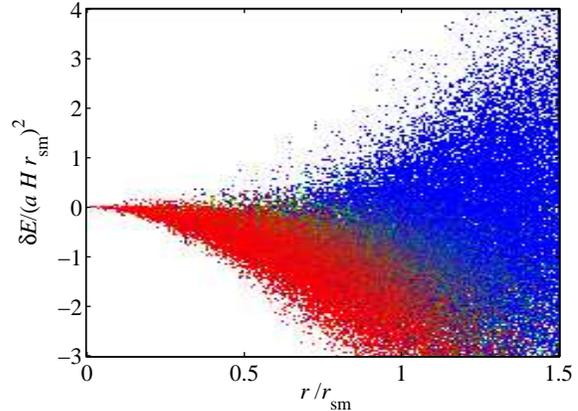}}
\caption{Binding energy versus initial radius for peaks that are
  accreted (red), that become subhalos (green), or that escape (blue).
\label{Ebnd}
}
\end{figure}

One Lagrangian quantity which correlates with survival is plotted in
Figure \ref{Ebnd}.  We have taken halos of mass 40,000-60,000
particles at $a=1$, nearly $M\simeq M_\star$, and identified
smaller-scale peaks within the spherical Lagrangian volumes centered
on the larger halos.  As shown in the figure, the relative binding
energy of the subpeaks roughly correlates with their escape
likelihood.  We have defined the binding energy $\delta E$ in the
following manner.  The gravitational potential near a peak may be
written
\begin{equation}
\phi({\bm x})\approx \phi_0 + {\bm x}\cdot\nabla\phi + 
\frac{1}{2}{\bm x}\cdot\nabla\nabla\phi\cdot{\bm x} + \ldots
\end{equation}
The first two terms do not affect the collapse of the peak, and so we
define the perturbed potential energy as the quadratic term,   
\begin{equation}
\delta\phi({\bm x})=\frac{1}{2}{\bm x}\cdot\nabla\nabla\phi\cdot{\bm x}
= \frac{3\Omega H_0^2}{4a}{\bm x}\cdot{\bf S}\cdot{\bm x}
\end{equation}
where we have used the Poisson equation to relate the strain tensor
${\bf S} = \nabla\nabla (\nabla^{-2}\delta)$ to the potential's
curvature tensor.  We can similarly write the kinetic energy near the
peak as 
\begin{eqnarray}
T({\bm x})&=&\frac{1}{2}|{\bm v}|^2=
\frac{1}{2}|aH({\bm x}+{\bm d}) + {\bm v}_p|^2 \nonumber \\
&\approx& \frac{1}{2}(aH|{\bm x}|)^2 + 2 (aH)^2 {\bm x}\cdot{\bm d} + \ldots
\end{eqnarray}
where we have used the Zeldovich approximation for $\Omega_m=1$
to relate the peculiar
velocity and comoving displacement, ${\bm v}_p=a H {\bm d}$.  The
unperturbed piece again does not affect the peak dynamics, and
so we write the perturbed kinetic energy as
\begin{equation}
\delta T = 2 (aH)^2 {\bm x}\cdot{\bm d} ,
\end{equation}
and the total perturbed energy becomes $\delta E = \delta\phi + \delta T$.  
In computing $\delta\phi$, we smooth the strain tensor ${\bf S}$ over
the large halo, whereas in $\delta T$ we use the difference in the
mean Lagrangian peculiar velocities, $\Delta{\bm v}_{\rm L}=
{\bm v}_{{\rm L}2} - {\bm v}_{{\rm L}1}$.   

As can be seen from Figure \ref{Ebnd}, the initial Lagrangian binding
energy does not solely determine the survival of subpeaks; the escape
threshold is radially dependent as well.  Without any physical
justification, we find that an approximate threshold
\begin{equation}
E^\prime \equiv \delta E + (a H)^2 [2 r^2 - r_{\rm sm}^2] = 0
\label{eqn:Ebnd}
\end{equation}
roughly separates accreted subpeaks from escaping subpeaks.

Our assumption, therefore, is that subpeaks inside of larger peaks can
escape accretion onto their hosts if the relative energy is large
enough, $E^\prime>0$.  Due to environmental effects, however, they are
unable to grow, and thereby produce the population of old, unbiased
low-mass halos.  In the next section, we test how well this simple
picture can reproduce our numerical bias results.

\section{A toy model for assembly bias}

\begin{figure*}
\centerline{
\includegraphics[width=0.3\textwidth]{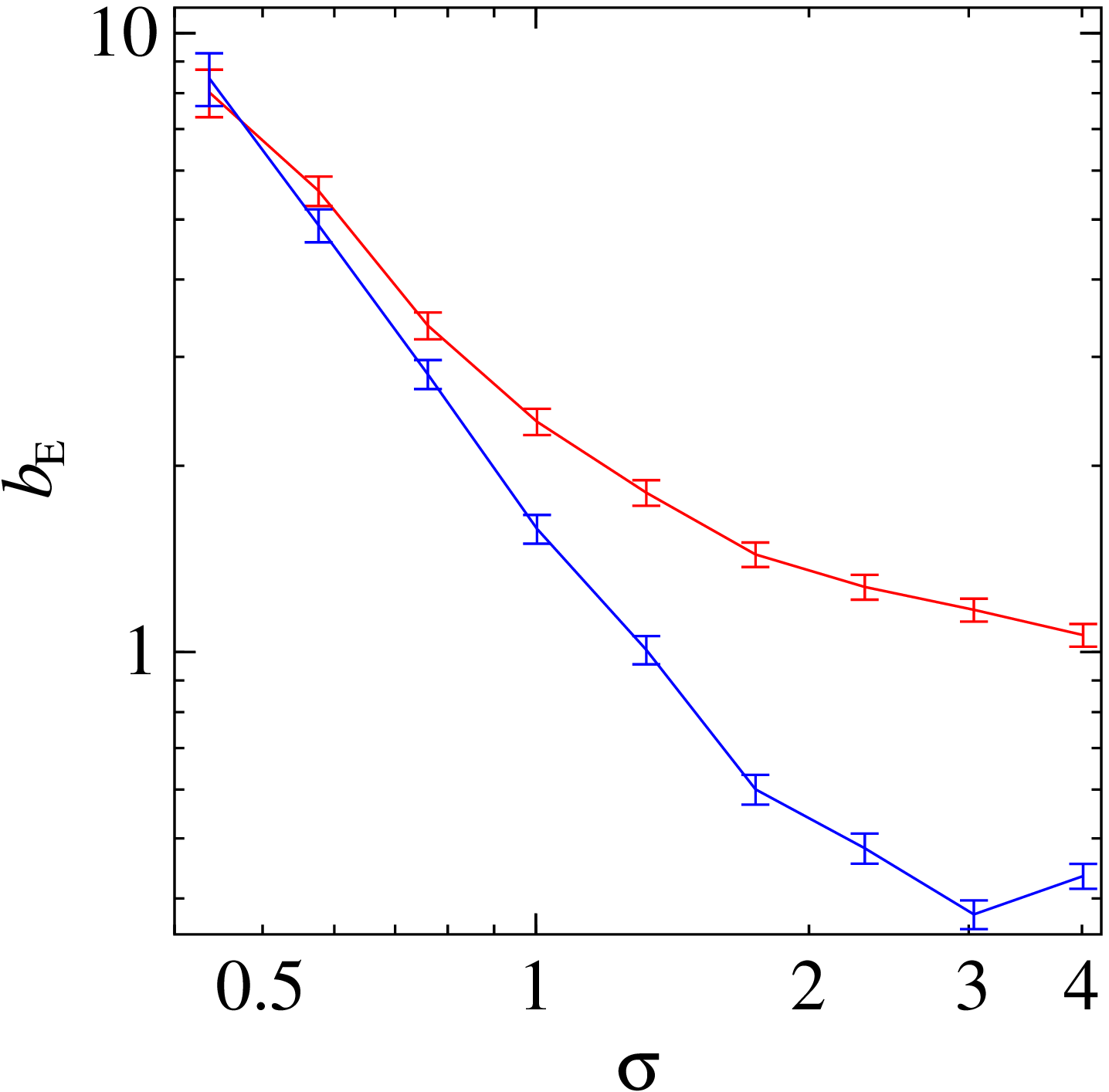} \qquad
\includegraphics[width=0.3\textwidth]{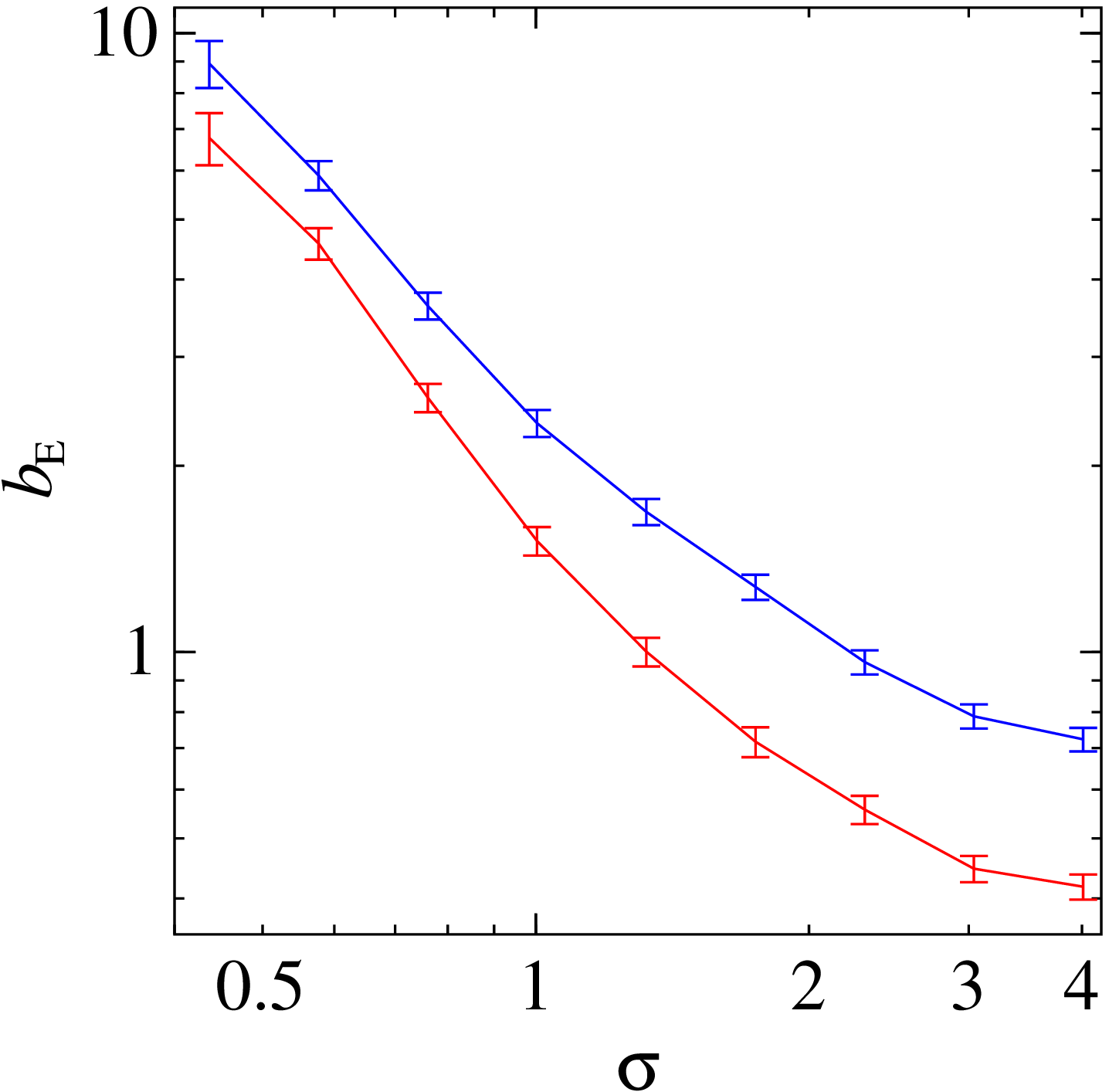} \qquad
\includegraphics[width=0.3\textwidth]{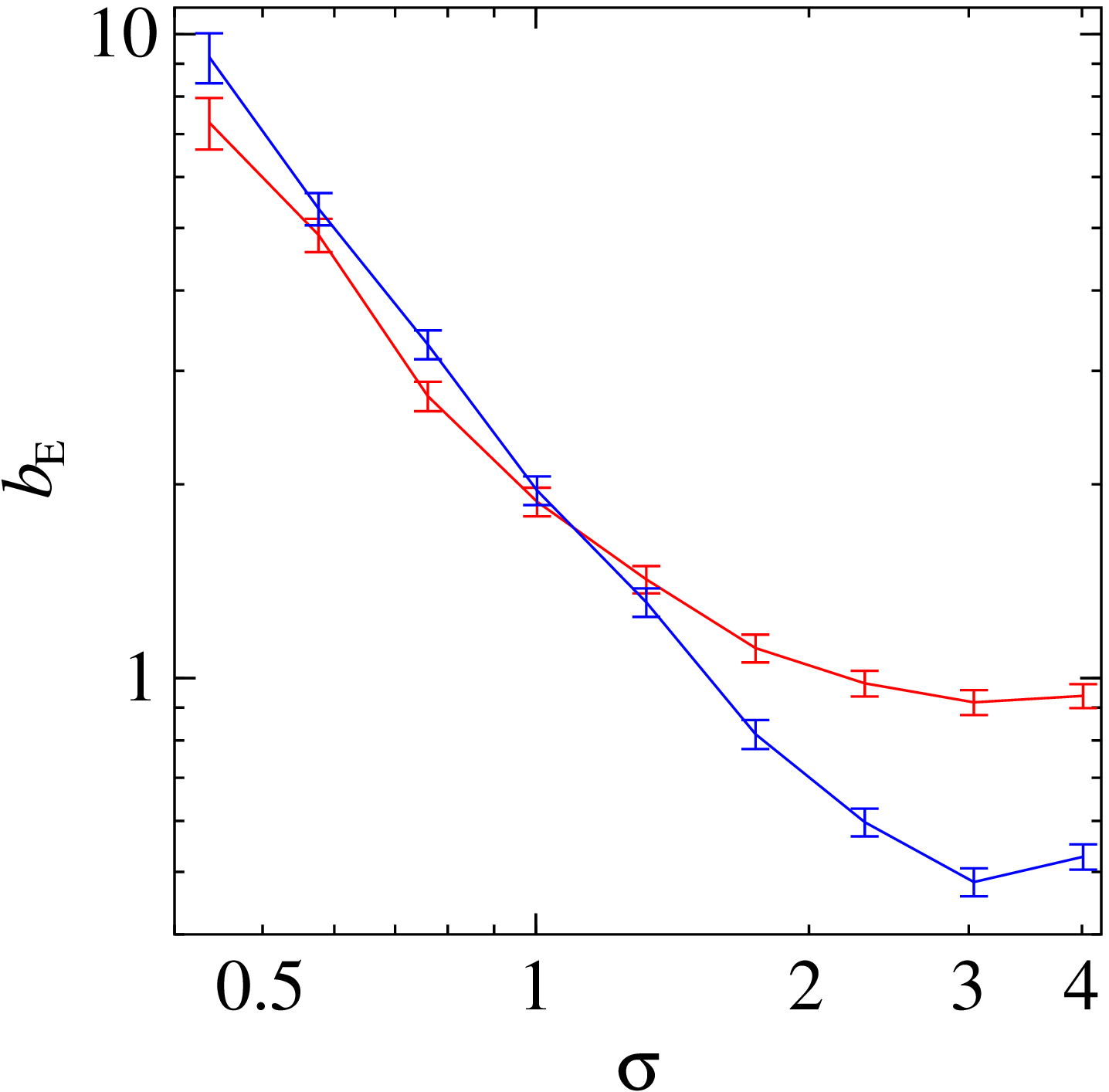}
}
\caption{Bias as a function of peak parameters.  Bias is plotted for
  halos in the range 200-300 particles at various redshifts.  In the
  left panel, the red and blue curves show bias for the upper and
  lower 20\% of peak height $\langle\delta\rangle/\delta_c$, where
  $\delta_c$ is the ellipsoidal collapse threshold calculated from each
  halo's local strain tensor.  In the middle panel, we split halos on
  peak curvature, $d\log\langle\delta\rangle/d\log M$.  In the right
  panel, we split halos on an average of these two
  quantities, $\langle\delta\rangle/\delta_c -
  d\log\langle\delta\rangle/d\log M$.  At high masses, the scatter in
  curvature dominates, while at low masses, the scatter in peak height
  dominates, leading to a transition in assembly bias at intermediate
  masses.  Formation time or concentration shows similar behavior.
\label{bias_sim}
}
\end{figure*}

In the previous two sections, we have discussed assembly bias in two
asymptotic regimes, $\sigma\ll 1$ and $\sigma \gg \delta_c$.  We found that
two distinct effects generated assembly bias in these two regimes.  At
intermediate masses, $\sigma \sim 1$, both effects can be significant,
and the overall bias of a sample of halos will involve a competition
between these two effects.  The transition between the high-mass
behavior and the low mass behavior will depend in detail upon the
precise selection criteria used to define the sample of halos.  For
example, we plot in Figure~\ref{bias_sim} the halo bias for halos
selected on peak height $\langle\delta\rangle$ and on peak curvature
$d\log\langle\delta\rangle/d\log M$.  Recall that these peak
properties both affect halo properties like age and concentration.  At
high masses, the scatter in peak height among peaks above threshold is
small, and so the scatter in peak curvature dominates the assembly
bias.  At low mass, the peak height is the dominant factor determining
halo age or concentration, causing a reversal in the sense of assembly
bias.  To illustrate, we plot in the right panel of
Fig.~\ref{bias_sim} the bias for halos selected on 
$\langle\delta\rangle/\delta_c - d\log\langle\delta\rangle/d\log M$.  
At high masses, the curvature term dominates, while at low masses, the
first term dominates.  Halo samples selected on formation time or
concentration show similar behavior.  At intermediate masses, both of
these effects are significant and compete with each other, and the
magnitude and sign of assembly bias will depend in detail upon the
precise age indicator that is employed.  For example, an age indicator
that correlates more strongly with peak height rather than peak
curvature will show a crossover at higher masses than the example
shown in Fig.~\ref{bias_sim}c.  This may explain why different
measures of halo age give different amounts of assembly bias when
applied to the same samples of halos \citep{LiMoGao08}.

Using our results on the assembly bias of low- and high-mass halos, we
have constructed a simple toy model for halo biasing.  We generate a
Gaussian random density field, smooth the field on multiple scales,
and starting from the largest smoothing scales, search for density
peaks exceeding the ellipsoidal collapse threshold \citep[hereafter
BM96]{BonMye96}. Isolated peaks above threshold are labeled as
collapsed halos, whereas peaks falling within 1.2 smoothing radii 
(c.f.\ Fig.~\ref{Ebnd}) of 
larger scale collapsed regions are only labeled as collapsed if their
binding energies relative to their larger hosts exceeds the threshold
found empirically from our N-body halos, i.e.\ Eqn.~(\ref{eqn:Ebnd}).

A key component of this calculation is the collapse threshold.
A comparison of the simple BM96 model with our N-body
results shows that the model works well at predicting collapse
thresholds for reasonably high peaks, $\sigma\lesssim 1$, however at
low peaks $\sigma \gg 1$, the model begins to break down.  Very
roughly, the ellipsoidal collapse threshold is larger than the
spherical collapse threshold by a ratio 
$\delta_{\rm ec}/\delta_{\rm sc}\sim (1-\frac{1}{2}\delta_{\rm sc}e_v)^{-1}$, 
where the ellipticity $e_v$ of the strain tensor is given by the
difference between its largest and smallest eigenvalues,
$e_v=(S_{33}-S_{11})/2\delta$.  The ellipticity and prolaticity are
random variables whose probability distribution is related to peak
height \citep[BM96]{BBKS}, but very roughly the mean ellipticity
scales as $\langle e_v\rangle \sim 1/(2\nu)$.  For $\nu<1$, the BM96
collapse threshold quickly grows, and for sufficiently large $e_v$ and
$-p_v$ the model predicts that collapse never occurs. However we find
many low mass halos with average peak heights $\langle\delta\rangle$
falling far below the collapse thresholds expected given their average
Lagrangian strain tensors $\langle{\bf S}\rangle$.  This is not
surprising.  As noted explicitly by BM96, their
homogeneous ellipsoid model assumes that the evolution of the external
tidal field acting upon the collapsing peak may be determined from the
strain tensor $\langle{\bf S}\rangle$ averaged over the local volume
of the peak.  While it is certainly true that the external tidal field
and local strain tensor are closely related at early times, in the
nonlinear regime the one-to-one correspondence between them breaks
down.  On scales where $\sigma\gg 1$, the external tidal field evolves
nonlinearly in a manner that cannot be predicted from purely local
measurements over the peak volume.  Therefore, we would not expect
the simple BM96 model to apply in this regime, as those authors note.

Nevertheless, we require some choice of collapse threshold for our toy
model.  At low masses, the lowest $\langle\delta\rangle$'s are roughly
$\sim 2.5$ (see Fig.~\ref{bias_lowm}), so we arbitrarily set a
threshold $\delta_c = {\rm min}(\delta_{\rm BM96},2.5)$.  In addition,
we also require that subpeaks collapsing inside of larger peaks must
complete their collapse sufficiently before the larger peak
collapses.  We can estimate the collapse redshift $z_c$ by
$1+z_c=\langle\delta\rangle/\delta_c$.  Writing $z_s$ and $z_b$ as the
collapse redshifts for the sub-peak and large peak, respectively, we
require $(1+z_s)\ge 2(1+z_b)$ for a subpeak to be labeled as a
collapsed halo. 

\begin{figure*}
\centerline{
\includegraphics[width=0.3\textwidth]{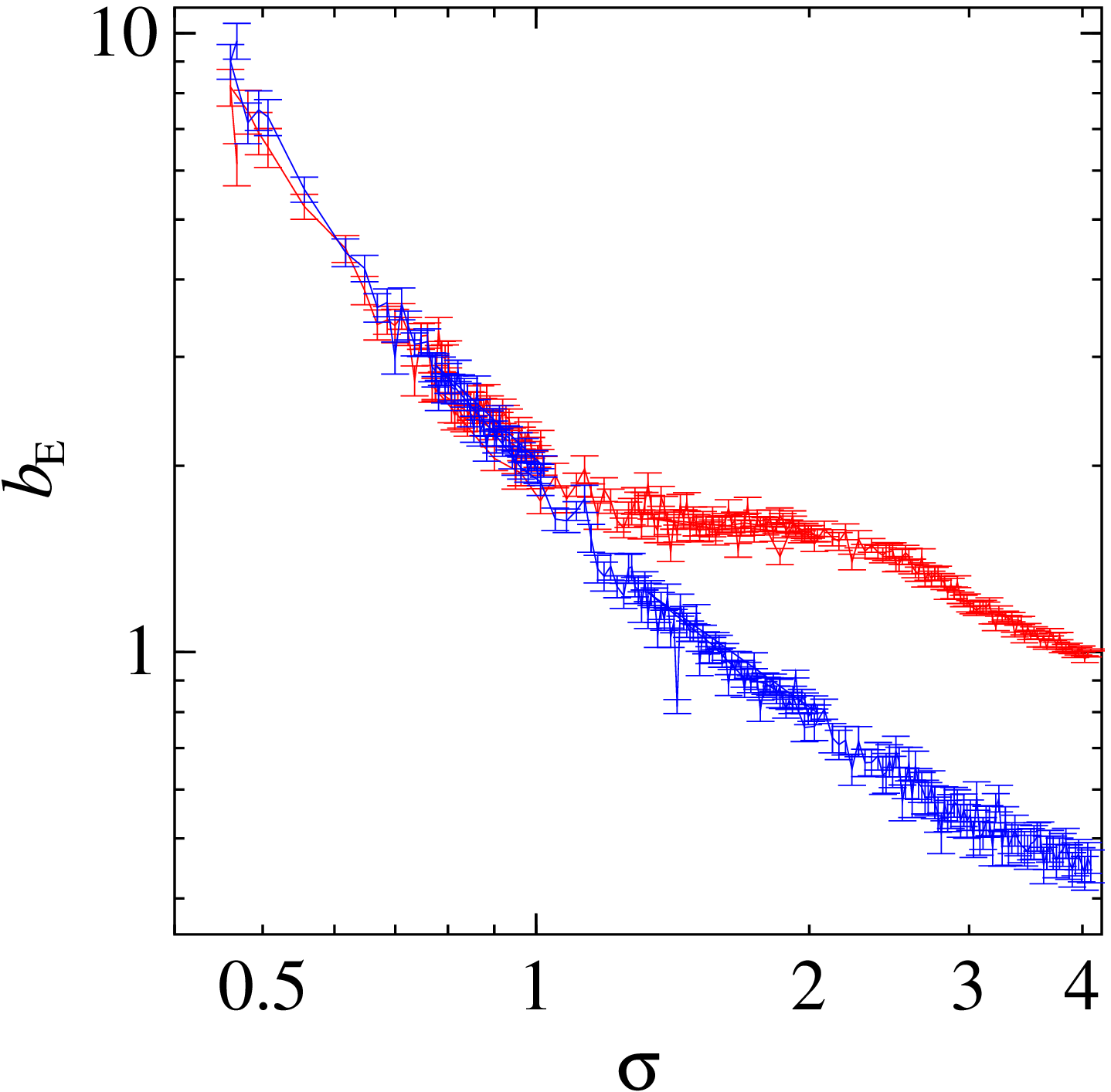} \qquad
\includegraphics[width=0.3\textwidth]{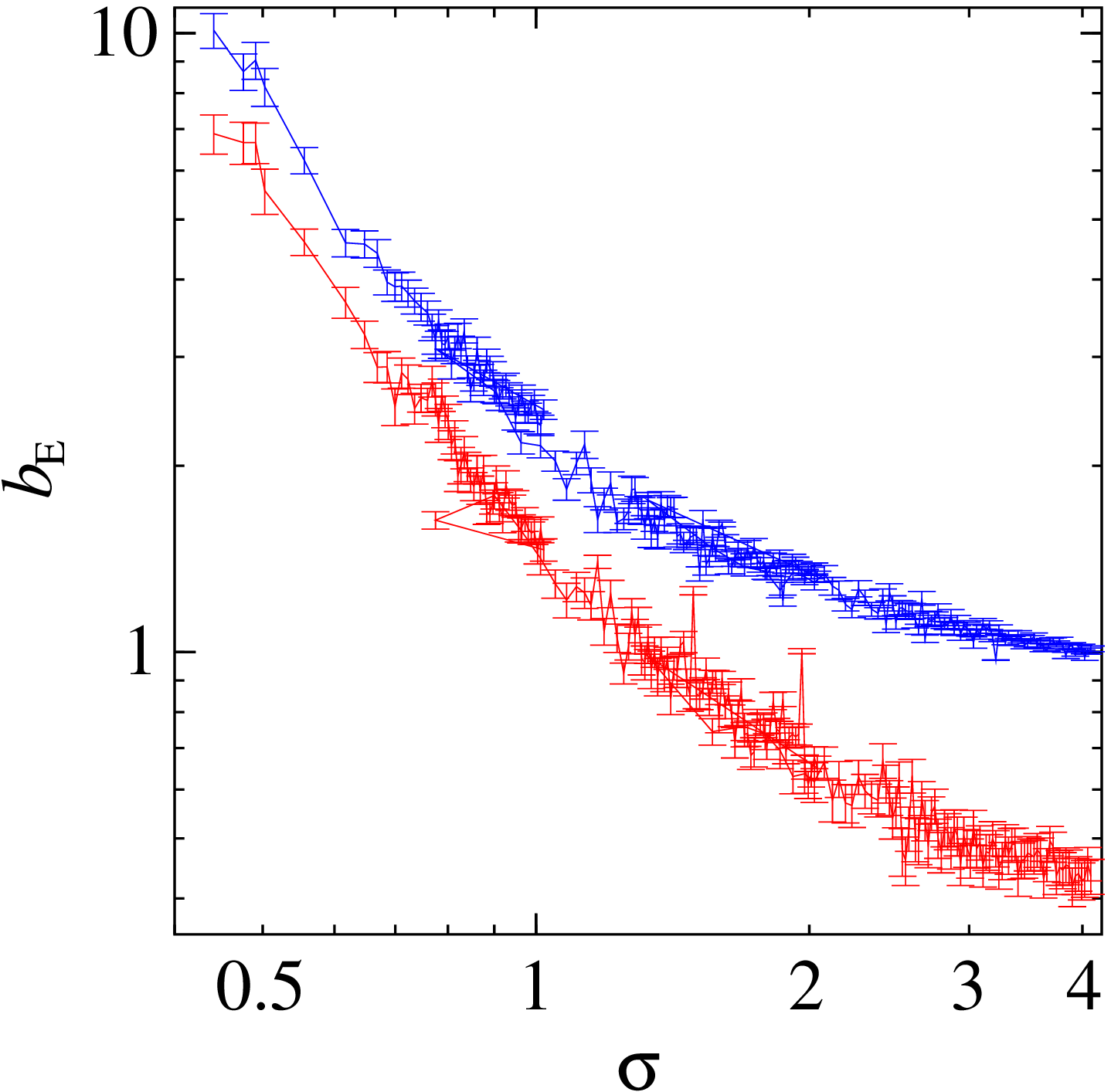} \qquad
\includegraphics[width=0.3\textwidth]{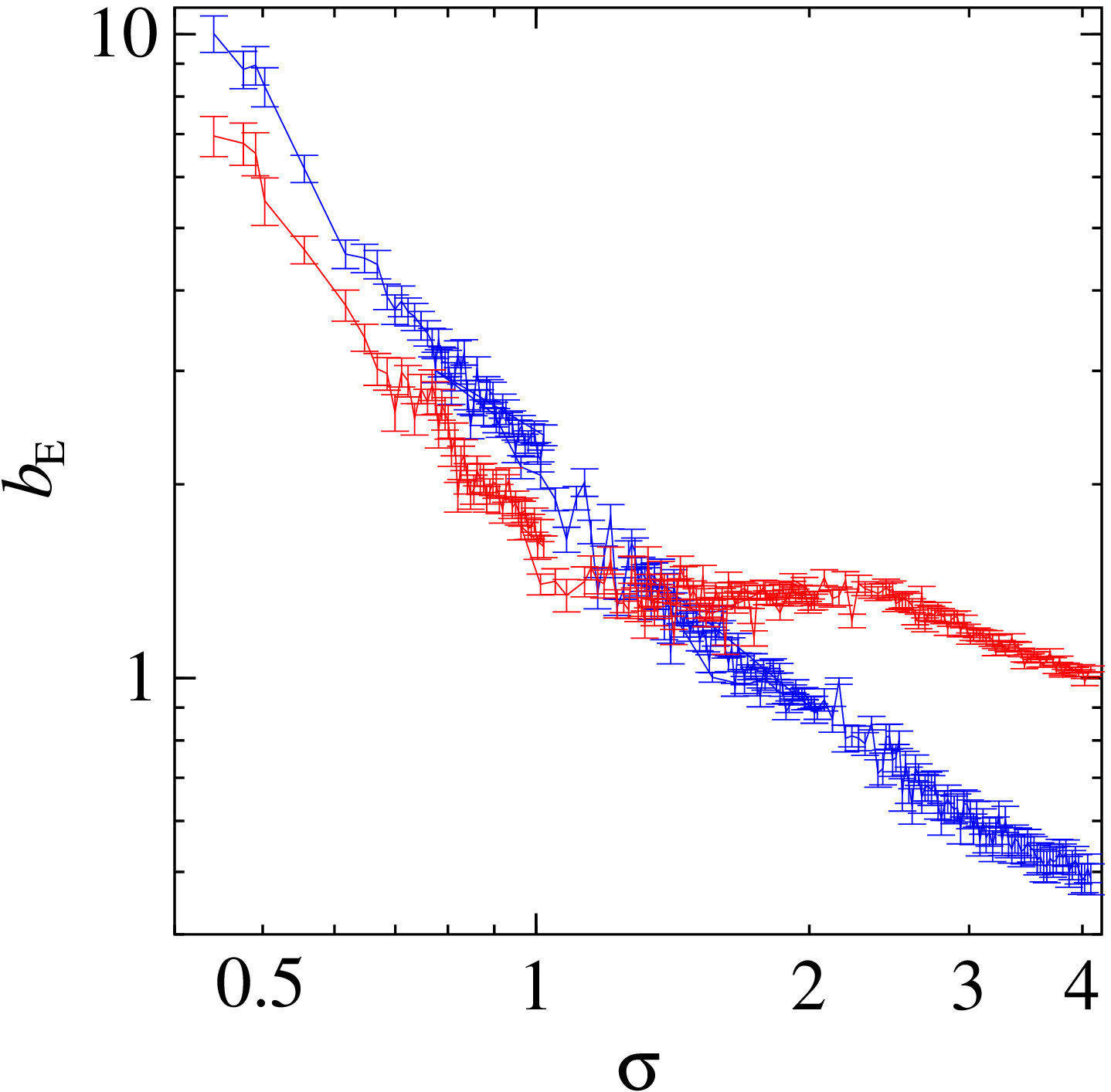}
}
\caption{Similar to Fig.~\ref{bias_sim}, but for peaks found in our
  toy model calculation rather than N-body halos.
\label{bias_toy}
}
\end{figure*}

Figure \ref{bias_toy} shows results of this simple calculation.  In
order to compare with Fig.~\ref{bias_sim}, we plot bias as a function
of the same peak parameters.  At high masses,
$\sigma\ll 1$, the bias matches Eqn.~\ref{bias_curv}, as it must.  At
low masses ($\sigma\gg 1$) the bias of isolated peaks asymptotes to
$b_{\rm E}\approx 0.4$, while the subpeaks become unbiased, 
$b_{\rm E}\to 1$.  These features appear fairly generic.  Behavior in
the transition region, $\sigma\sim 1-2$, depends upon details of the
model.  In particular, the bias of the old halos in the intermediate
mass regime depends upon the fraction of halos that are subpeaks, and
varying parameters in the model changes this fraction.

Overall, however, we find reasonable agreement between the toy model
and our N-body simulation in the regimes where assembly bias is
significant, for $M\gg M_\star$ and $M\ll M_\star$.  The point of this
exercise is not to claim that we have solved the halo formation
problem, but merely to illustrate that the assembly bias in these
regimes is generic.  The assembly bias at high masses is an
unavoidable consequence of Gaussian statistics, and therefore any halo
formation model which relates initial density peaks to halos will also
find similar levels of assembly bias, depending somewhat on the peak
definition.  The behavior at low masses may appear more model
dependent, however we do not believe our results are specific to our
particular mechanism for production of non-accreting halos. Rather, we
expect that any mechanism which produces roughly the correct
fraction of non-growing, low-mass halos will also match the
assembly bias found in N-body simulations.

\section{Discussion \& Conclusions}

We have investigated halo assembly bias in
hierarchical structure formation.  We find that assembly bias is
significant both at high masses ($M\gg M_\star$) and low masses 
($M\ll M_\star$).  Two different competing effects generate assembly
bias in these two regimes.  At high masses, peaks of low curvature are
more strongly clustered than peaks of large curvature, in a manner
entirely consistent with the statistics of peaks of random Gaussian
fields.  We provide a simple expression for the bias in the high-mass
regime which accounts for assembly bias and reduces to the
Press-Schechter bias when marginalized over all parameters excluding
mass (note that fitting formulae for bias calibrated at lower mass,
such as those proposed by \citet{SheTor99} or \citet{SelWar04}, fail
in the high mass regime \citep{CohWhi07}). 
At low masses, assembly bias appears largely driven by the
formation of a non-accreting sub-population of low-mass halos in the
vicinity of larger halos.  Because the non-accreting halos become
unbiased, they are more strongly clustered than accreting halos of
similar mass, which are anti-biased.  
We found that a simple toy model is able to reproduce our
numerical bias results in the asymptotically large and small mass
regimes.  At intermediate masses ($M\sim M_\star$), both of the above
effects become important, and the assembly bias for a sample of halos
will involve a weighted average of both effects.

The application of these results to observed objects like galaxies or
quasars is not straightforward, because it will depend upon the halo
occupation distribution (HOD) of those objects.  The very simplest HOD
models assume that the galaxy content depends only upon halo mass, and
not upon assembly history, whereas semi-analytic models used in
conjunction with N-body simulations tend to find a strong dependence
of galaxy properties on assembly history at fixed mass 
\citep[e.g.]{Millenium,Cro07}.  If the former assumption is
valid, then assembly bias will not be important for galaxy
clustering.  Perhaps surprisingly, \citet{Tin07} have recently found
that properties of observed SDSS galaxies do {\em not} appear to
correlate significantly with local environment, once the dependence on
host halo mass has been accounted for, suggesting that assembly bias
may indeed be unimportant for modeling galaxy clustering.

Besides the clustering of halos, our results also have implications
for our understanding of halo assembly.  In particular, we have found
that massive halos are very well described by spherical collapse.
The average pre-collapse Lagrangian volumes occupied by the most
massive halos are approximately spherical regions of size 
$r \sim r_{\rm TH}=(3M/4\pi\rho)^{1/3}$.  
However, $r_{\rm TH}$ is not a sharp boundary in Lagrangian space:
there is significant mass outside $r>r_{\rm TH}$ at early times which
is found inside the virial radius at late times, and similarly there
is considerable mass at $r<r_{\rm TH}$ which avoids accretion at late
times.  Indeed, we might speculate that the post-collapse radial
distribution of the matter inside and around a dark matter halo is
simply related to the pre-collapse profile of local density and
potential.  Since both of these quantities are described by
well-understood Gaussian statistics, this raises the intriguing
possibility of understanding the internal structure of virialized
objects from first principles.  This topic is the subject of further
study.

\acknowledgments

We thank Andrey Kravtsov and Simon White for several helpful
discussions.  We also thank Joanne Cohn, Andrey Kravtsov and Andrew
Wetzel for comments on an earlier version of this paper.
All simulations were performed on CITA's Sunnyvale
cluster, funded by the Canada Foundation for Innovation and the
Ontario Research Fund for Research Infrastructure.  N.D. was supported
by the Natural Sciences and Engineering Research Council of Canada
(NSERC).  M.W. was supported in part by NASA.

\bibliographystyle{apj}
\end{document}